\documentclass[prd,twocolumn,nofootinbib,superscriptaddress,amssymb,amsfonts,amsmath,preprintnumbers]{revtex4-2}
\usepackage{amsmath}
\usepackage{dsfont}
\usepackage{mathrsfs}
\usepackage{scalerel}
\usepackage[colorlinks=true,linktoc=page,linkcolor=purple,citecolor=teal,urlcolor=magenta]{hyperref}
\usepackage{url}
\usepackage{caption}
\usepackage{subcaption}
\usepackage{graphicx,tabularx,tabularray,diagbox}
\usepackage[compat=1.1.0]{tikz-feynman}
\usepackage{xcolor}
\definecolor{dgreen}{rgb}{.0,.6,.0}
\definecolor{lime}{HTML}{A6CE39}
\definecolor{lg}{RGB}{220,220,220} 
\usepackage{ragged2e}
\usepackage{natbib}
\usepackage{cancel}
\makeatletter
\g@addto@macro\bfseries{\boldmath}
\makeatother
\DeclareRobustCommand{\orcidicon}{\hspace{-2.1mm}
\begin{tikzpicture}
\draw[lime,fill=lime] (0,0.0) circle [radius=0.13] node[white] {{\fontfamily{qag}\selectfont \tiny ID}}; \draw[white,fill=white] (-0.0525,0.095) circle [radius=0.007]; 
\end{tikzpicture} \hspace{-3.7mm} }
\foreach \x in {A, ..., Z} {\expandafter\xdef\csname orcid\x\endcsname{\noexpand\href{https://orcid.org/\csname orcidauthor\x\endcsname} {\noexpand\orcidicon}}}

\newcommand{\Eprint}[1]{\href{#1}}

\newcommand\wh[1]{\hstretch{2}{\hat{\hstretch{.5}{#1\mkern1mu}}}\mkern-1mu}

\newenvironment{Eqnarray}%
     {\arraycolsep 0.14em\begin{eqnarray}}{\end{eqnarray}}
\def\beq{\begin{Eqnarray}}
\def\eeq{\end{Eqnarray}}
\def\beqa{\begin{Eqnarray}}
\def\eeqa{\end{Eqnarray}}
\def\bea{\begin{Eqnarray}}
\def\eea{\end{Eqnarray}}
\def\eq#1{Eq.~(\ref{#1})}
\def\Eq#1{Eq.~(\ref{#1})}

\def\eqs#1#2{Eqs.~(\ref{#1}) and (\ref{#2})}

\def\eqst#1#2{Eqs.~(\ref{#1})--(\ref{#2})}

\def\cbma{c_{\beta-\alpha}}
\def\sbma{s_{\beta-\alpha}}
\def\sbmaii{s^2_{\beta-\alpha}}
\def\cbmaii{c^2_{\beta-\alpha}}
\def\sbmaiii{s^3_{\beta-\alpha}}
\def\cbmaiii{c^3_{\beta-\alpha}}

\def\ifmath#1{\relax\ifmmode #1\else $#1$\fi}
\def\lsub#1{\ifmath{_{\lower2.5pt\hbox{$\scriptstyle #1$}}}}
\def\lsup#1{^{\lower 2pt\hbox{$\scriptstyle#1$}}}
\let\Re\relax
\let\Im\relax
\DeclareMathOperator{\Re}{Re}
\DeclareMathOperator{\Im}{Im}
\def\vev#1{\left\langle #1\right\rangle}
\def\half{\tfrac12}

\def\nn{\nonumber}
\makeatletter
\let\save@mathaccent\mathaccent
\newcommand*\if@single[3]{%
  \setbox0\hbox{${\mathaccent"0362{#1}}^H$}%
  \setbox2\hbox{${\mathaccent"0362{\kern0pt#1}}^H$}%
  \ifdim\ht0=\ht2 #3\else #2\fi
  }
\newcommand*\rel@kern[1]{\kern#1\dimexpr\macc@kerna}
\newcommand*\widebar[1]{\@ifnextchar^{{\wide@bar{#1}{0}}}{\wide@bar{#1}{1}}}
\newcommand*\wide@bar[2]{\if@single{#1}{\wide@bar@{#1}{#2}{1}}{\wide@bar@{#1}{#2}{2}}}
\newcommand*\wide@bar@[3]{%
  \begingroup
  \def\mathaccent##1##2{%
    \let\mathaccent\save@mathaccent
    \if#32 \let\macc@nucleus\first@char \fi
    \setbox\z@\hbox{$\macc@style{\macc@nucleus}_{}$}%
    \setbox\tw@\hbox{$\macc@style{\macc@nucleus}{}_{}$}%
    \dimen@\wd\tw@
    \advance\dimen@-\wd\z@
    \divide\dimen@ 3
    \@tempdima\wd\tw@
    \advance\@tempdima-\scriptspace
    \divide\@tempdima 10
    \advance\dimen@-\@tempdima
    \ifdim\dimen@>\z@ \dimen@0pt\fi
    \rel@kern{0.6}\kern-\dimen@
    \if#31
      \overline{\rel@kern{-0.6}\kern\dimen@\macc@nucleus\rel@kern{0.4}\kern\dimen@}%
      \advance\dimen@0.4\dimexpr\macc@kerna
      \let\final@kern#2%
      \ifdim\dimen@<\z@ \let\final@kern1\fi
      \if\final@kern1 \kern-\dimen@\fi
    \else
      \overline{\rel@kern{-0.6}\kern\dimen@#1}%
    \fi
  }%
  \macc@depth\@ne
  \let\math@bgroup\@empty \let\math@egroup\macc@set@skewchar
  \mathsurround\z@ \frozen@everymath{\mathgroup\macc@group\relax}%
  \macc@set@skewchar\relax
  \let\mathaccentV\macc@nested@a
  \if#31
    \macc@nested@a\relax111{#1}%
  \else
    \def\gobble@till@marker##1\endmarker{}%
    \futurelet\first@char\gobble@till@marker#1\endmarker
    \ifcat\noexpand\first@char A\else
      \def\first@char{}%
    \fi
    \macc@nested@a\relax111{\first@char}%
  \fi
  \endgroup
}
\makeatother

\begin{document}

\preprint{ZU-TH 89/25}
\title{
Correlating Resonant Di-Higgs and Tri-Higgs Production to $H\to VV$ in the 2HDM}

\author{Guglielmo Coloretti\orcidA{}}
\email{guglielmo.coloretti@bo.infn.it}
\affiliation{INFN, Sezione di Bologna, Via Irnerio 46, 40126 Bologna, Italy}

\author{Andreas Crivellin\orcidB{}}
\email{andreas.crivellin@icrea.cat}
\affiliation{Universitat Autònoma de Barcelona, 08193 Bellaterra, Barcelona.}
\affiliation{ICREA, Instituci\'o Catalana de Recerca i Estudis Avan\c{c}ats,
Passeig de Llu\'{\i}s Companys 23, 08010 Barcelona, Spain.}

\author{Howard E. Haber\orcidD{\,\,}}
\email{haber@scipp.ucsc.edu}
\affiliation{Santa Cruz Institute for Particle Physics, University of California, 1156 High Street, Santa Cruz, CA 95064, USA}

\begin{abstract}
The observation of resonant di-Higgs production, which would strongly suggest the existence of a new heavy neutral scalar $H$, has been searched for extensively at the LHC. In the two-Higgs doublet model (2HDM) with $m_H\gg m_h$, where $h$ is the Higgs boson of mass 125 GeV observed at the LHC, we show that a direct correlation emerges between Br$(H\to hh)$ and Br$(H\to VV)$, with $V=Z,W$, which depends only on $m_H$ (and $m_V$). In particular, for heavy scalar masses between 500\,GeV and 1\,TeV, we find that Br($H\to hh$)/ Br($H\to ZZ)\approx 9.4\pm 0.25$. 
Moreover, $H\to hh$ is a dominant decay mode over a significant region of the parameter space and serves as the primary probe for a heavy scalar resonance at current and future hadron colliders.
The origin of these predictions is most transparent in the Higgs basis, where the term in the scalar potential proportional to 
$\mathcal H_1^\dagger  \mathcal H_1  \mathcal H_1^\dagger  \mathcal H_2$ (and its hermitian conjugate) generates the leading contributions to the
$Hhh$ and $Hhhh$ couplings in the decoupling limit of the 2HDM. Additionally, the latter coupling governs the resonant prompt tri-Higgs production via $H\to hhh$, which is also directly correlated to $H\to hh$ (and $H\to VV$), and can yield rates large enough to be measured at the High-Luminosity LHC. 
\end{abstract}
\maketitle

\section{Introduction}

With the discovery of the Higgs boson at the Large Hadron Collider (LHC) at CERN~\cite{Aad:2012tfa,Chatrchyan:2012ufa} (denoted by $h$ with $m_h\simeq 125$\,GeV), the spectrum of fundamental particles of the Standard Model (SM) is now completely observed. Moreover, the properties of the Higgs boson match the corresponding predictions of the SM within the experimental accuracy of the measurements~\cite{ATLAS:2022vkf,ATLAS:2025qxq,CMS:2022dwd,CMS:2026nce}.

Nevertheless, it is striking that the scalar sector employed by the SM is minimal, in contrast to the nonminimal structure of the fermionic sector that exhibits a replication of quark and lepton generations.  One might therefore anticipate a nonminimal scalar sector in which the Higgs doublet field of the SM is replicated as well. Indeed, there are hints for the existence of new particles in general, and new Higgs bosons in particular, with masses of order 1 TeV and below~\cite{Crivellin:2023zui}. Importantly, even if current hints for new physics at the LHC turn out to be statistical fluctuations, there is still ample motivation to study the phenomenological implications of extensions of the SM scalar sector.  

A smoking-gun signature of extended Higgs sectors containing new scalars heavier than $2m_h$ is resonant di-Higgs production. This process has been studied in the context of multiple models (see Refs.~\cite{Muhlleitner:2017dkd,Abouabid:2021yvw} for a review), including the scalar singlet-extended SM~\cite{Barger:2008jx,Costa:2015llh,Dawson:2017jja,No:2018fev}, the 2HDM~\cite{Bernon:2015qea,Grzadkowski:2018ohf,Kling:2020hmi,Arco:2022lai,Haber:2022gsn} and more involved setups~\cite{Maniatis:2009re,Hartling:2014zca,Zhang:2015mnh}.  Indeed, resonant di-Higgs production is searched for extensively at the LHC in multiple channels~\cite{ATLAS:2019qdc,ATLAS:2021ifb,ATLAS:2022hwc,ATLAS:2022xzm,ATLAS:2022fpx,ATLAS:2024lsk,CMS:2018ipl,CMS:2022kdx,CMS:2023boe,CMS:2026mwf,CMS:2025tqi}, with a recent combination resulting in impressively strong limits on the corresponding cross section~\cite{ATLAS:2023vdy}. In the event of a future observation of resonant di-Higgs production, it is imperative to develop predictions that can distinguish among the various models.

In this article, we will examine resonant di-Higgs production in the context of the two-Higgs-doublet model (2HDM), which was originally introduced in Ref.~\cite{Lee:1973iz}, and has been extensively developed and analyzed in the literature.\footnote{For comprehensive reviews of the 2HDM, see, e.g., Refs.~\cite{Gunion:1989we,Branco:2011iw,Ivanov:2017dad} and Chapter 6.2 of Ref.~\cite{Dreiner:2023yus}.}
In light of the LHC Higgs data, which imply that the properties of $h$ are approximately those of the SM Higgs boson~\cite{ATLAS:2022vkf,ATLAS:2025qxq,CMS:2022dwd,CMS:2026nce},
the parameters of the 2HDM must be consistent with the so-called Higgs alignment limit~\cite{Ginzburg:2001wj,Gunion:2002zf,Craig:2012vn,Craig:2013hca,Asner:2013psa,Carena:2013ooa,Haber:2013mia,BhupalDev:2014bir}.
The Higgs alignment condition is naturally satisfied in the decoupling limit of the 2HDM~\cite{Haber:1989xc,Gunion:2002zf}, where all the new scalar states are significantly heavier than $h$, under the assumption of perturbative scalar self-couplings. In our analysis, we shall simplify the presentation by first neglecting possible new CP-violating effects in the scalar sector, with a focus on the production and decay of the heavy CP-even neutral scalar, denoted by~$H$.  A brief treatment of the CP-violating case will be given at the end of this work.

In the decoupling limit (where $m_H\gg m_h$), we will show that the 2HDM exhibits direct correlations of resonant di-Higgs production with $H\to VV$  ($V=W$ or $Z$) that depend only on $m_H$ (and $m_V$). This feature can be most easily understood in the Higgs basis~\cite{Georgi:1978ri,Lavoura:1994yu,Lavoura:1994fv,Botella:1994cs,Branco:1999fs,Davidson:2005cw}, where the Higgs vacuum expectation value (vev) resides entirely in one of the two Higgs doublet fields.  In the decoupling limit, the leading contribution to the $Hhh$ coupling arises from a single term in the Higgs basis scalar potential, which is proportional to a parameter denoted by $Z_6$. 
Moreover, $Z_6$ also gives rise to the leading effect in resonant prompt tri-Higgs production ($pp\to H\to 3h$), where the first measurement attempts at the LHC have recently been made~\cite{CMS:2025jkb,CMS:2025gos,ATLAS:2024xcs}.  The cross sections for resonant di-Higgs and tri-Higgs production may be large enough to be observable at the High-Luminosity LHC (HL-LHC)~\cite{Apollinari:2017lan,Cepeda:2019klc} and FCC-hh~\cite{Mangano:2017tke,FCC:2018vvp}. 

\section{2HDM Formalism}
\label{formalism}
The 2HDM employs two SU(2)$_L$ complex scalar doublet fields $\Phi_1$ and $\Phi_2$, each with hypercharge $Y=\half$. In general, the neutral components of the two doublet fields acquire vevs, $\vev{\Phi_i^0}=v_i/\sqrt{2}$, where $v_1\equiv v c_\beta$ and $v_2 \equiv v s_\beta e^{i\xi}$, with
$v\simeq 246$\,GeV in a convention where $0\leq\beta\leq\half\pi$ and $0\leq\xi<2\pi$,
resulting in the breaking of the SU(2)$_L\times$U(1)$_Y$ gauge group to U(1)$_{\rm EM}$.\footnote{We have employed the notation where $s_\beta\equiv\sin\beta$ and $c_\beta\equiv\cos\beta$.  Without loss of generality, one is free to apply a suitable U(1)$_Y$ transformation such that $v$ is real and nonnegative.}  

Without further assumptions, the choice of $\Phi_1$--$\Phi_2$ basis is arbitrary.  It is therefore convenient to introduce the Higgs basis fields 
$\mathcal{H}_1$ and $\mathcal{H}_2$ (following Ref.~\cite{Boto:2020wyf}):
\beqa 
 \mathcal{H}_1&=&\begin{pmatrix}\mathcal{H}_1^+\\[4pt] \mathcal{H}_1^0\end{pmatrix}\equiv c_\beta\Phi_1+s_\beta e^{-i\xi}\Phi_2\,,\label{hbasis1}\\
\mathcal{H}_2&=&\begin{pmatrix}\mathcal{H}_2^+\\[4pt] \mathcal{H}_2^0\end{pmatrix}=e^{i\eta}\bigl(-s_\beta e^{i\xi}\Phi_1+c_\beta \Phi_2\bigr)\,,  
\label{hbasis2}
\eeq
where the scalar potential is minimized so that only one of the scalar doublets acquires a nonzero vev (i.e., $\vev{\mathcal{H}_1^0}=v/\sqrt{2}$ and $\vev{\mathcal{H}_2^0}=0$).
The complex phase factor $e^{i\eta}$ accounts for the nonuniqueness 
of the Higgs basis~\cite{Boto:2020wyf}, since one is always free to rephase the Higgs doublet field whose vacuum expectation value vanishes. 

In the Higgs basis, the most general renormalizable, SU(2)$_L\times$U(1)$_Y$ invariant scalar potential is 
given by
 \beqa
 \hspace{-0.03in}\mathcal{V}&=& Y_1 \mathcal{H}_1^\dagger \mathcal{H}_1+ Y_2 \mathcal{H}_2^\dagger \mathcal{H}_2 +[Y_3 e^{-i\eta}
\mathcal{H}_1^\dagger \mathcal{H}_2+{\rm h.c.}] \nn \\[5pt]
&& +\,\half Z_1(\mathcal{H}_1^\dagger \mathcal{H}_1)^2+\half Z_2(\mathcal{H}_2^\dagger \mathcal{H}_2)^2
+Z_3(\mathcal{H}_1^\dagger \mathcal{H}_1)(\mathcal{H}_2^\dagger \mathcal{H}_2) \nn \\[3pt]
&&
+Z_4(\mathcal{H}_1^\dagger \mathcal{H}_2)(\mathcal{H}_2^\dagger \mathcal{H}_1)
+\left\{\half Z_5  e^{-2i\eta}(\mathcal{H}_1^\dagger \mathcal{H}_2)^2\right. \nn \\
&&   \left. \,\,\,\,+\big[Z_6  e^{-i\eta}\mathcal{H}_1^\dagger
\mathcal{H}_1 +Z_7 e^{-i\eta} \mathcal{H}_2^\dagger \mathcal{H}_2\big] \mathcal{H}_1^\dagger \mathcal{H}_2+{\rm
h.c.}\right\},
\label{higgsbasispot}
\eeqa
where $Y_1$, $Y_2$, and $Z_1,\ldots,Z_4$ are real, whereas $Y_3$, $Z_5$, $Z_6$, and $Z_7$ are potentially complex parameters. The quantities $Y_1$ and $Y_3$ are fixed by the scalar potential minimization conditions, 
\beq \label{hbasisminconds}
Y_1=-\half Z_1 v^2\,,\qquad\quad Y_3=-\half Z_6 v^2\,.
\eeq 

The Higgs basis fields couple to up-type and down-type fermions via the Yukawa coupling Lagrangian,
\beqa
-\mathscr{L}_Y&=& \overline{Q}(\boldsymbol{\wh{\kappa}_{U}} \widetilde{\mathcal{H}}_1
 +  \boldsymbol{\wh{\rho}_{U}}  \widetilde{\mathcal{H}}_2\,) U  +  \overline{Q}(\boldsymbol{\wh{\kappa}_D^{\dagger}} {\mathcal{H}}_1
 +   \boldsymbol{\wh{\rho}_D^{\dagger}} {\mathcal{H}}_2\,) D\nn \\
&& + \overline{L}( \boldsymbol{\wh{\kappa}_E^{\dagger}}{\mathcal{H}}_1
 +   \boldsymbol{\wh{\rho}_E^{\dagger}} {\mathcal{H}}_2\,) E
+{\rm h.c.},\label{yukawas}
\eeqa
where $\widetilde{\mathcal{H}}_k\equiv i\sigma_2\mathcal{H}^{*}_k$ ($k=1,2$),
and $\boldsymbol{\wh{\kappa}_{F}}$ and $\boldsymbol{\wh{\rho}_{F}}$ 
are~$3\times 3$ Yukawa matrices in flavor space ($F=U,D,E$) that~are invariant 
with respect to scalar field basis transformations.
The quark and lepton fields appear in left-handed 
SU(2)$_L$ doublets $Q$ and $L$, and right-handed SU(2)$_L$ singlets 
$U$, $D$, and $E$ (with the generation indices suppressed).
When the fermion mass matrices \mbox{$\boldsymbol{\wh{M}_F}\equiv v\boldsymbol{\wh{\kappa}_F}/\sqrt{2}$,} 
are diagonalized via a singular value decomposition~\cite{Haber:2006ue}, the corresponding transformed $\boldsymbol{\wh{\rho}_F}$ matrices are, in general, complex and nondiagonal, which can yield dangerously large tree-level flavor-changing neutral currents (FCNCs) mediated by neutral scalars (e.g., see Ref.~\cite{Crivellin:2013wna} for a detailed analysis).

Here, we shall avoid off-diagonal neutral scalar couplings to fermions by imposing flavor alignment~\cite {Pich:2009sp,Pich:2010ic,Eberhardt:2020dat,Choi:2020uum,Lee:2021oaj,Serodio:2011hg,Cree:2011uy,Connell:2023jqq,Karan:2023kyj}.  For this, we define (potentially complex) flavor-alignment parameters $a_F$ in \eq{yukawas} via
\beq \label{alignedparms}
\boldsymbol{\wh{\rho}_F}=\frac{\sqrt{2}}{v} a_F \boldsymbol{\wh{M}_F}\,,\qquad\quad \text{for $F=U,D,E$},
\eeq
where the $a_F$ are invariant with respect to 
a change of the scalar field basis. As shown in Ref.~\cite{Ferreira:2010xe}, \eq{alignedparms} is not stable under renormalization group running except in special cases corresponding to the existence of a $\mathbb{Z}_2$ symmetry that is manifestly realized in a particular scalar field basis.  This is the case of natural flavor conservation~\cite{Glashow:1976nt,Paschos:1976ay}, which is often assumed in
2HDM phenomenological studies. However, to be more general, we do not make this assumption here, as phenomenological considerations do not require it.\footnote{Flavor-aligned extended Higgs sectors can arise naturally from symmetries of ultraviolet completions of low-energy effective theories of flavor as shown in Refs.~\cite{Knapen:2015hia,Egana-Ugrinovic:2018znw,Egana-Ugrinovic:2019dqu}. In such models, departures from exact flavor alignment due to renormalization group running down to the electroweak scale are typically small enough~\cite{Braeuninger:2010td,Gori:2017qwg} to be consistent with all known FCNC bounds.}

In light of the form of the scalar potential [\eq{higgsbasispot}] and the flavor alignment condition [\eq{alignedparms}], the scalar sector of the 2HDM yields potentially new sources of CP violation beyond the phase in the Cabibbo-Kobayashi-Maskawa (CKM) mixing matrix. To simplify the analysis presented in this work, we assume that all new sources of CP violation are absent. In particular, we shall assume that the flavor alignment parameters $a_F$ defined in \eq{alignedparms} are real. In addition, we shall also
assume the existence of a real $\Phi_1$--$\Phi_2$ basis in which all the parameters of the scalar potential are real, and the corresponding scalar field vevs are real (i.e., CP is not spontaneously broken by the scalar potential).  It then follows that a real Higgs basis exists where the potentially complex scalar potential parameters
$Y_3$, $Z_5$, $Z_6$, and $Z_7$ [cf.~\eq{higgsbasispot}]
are all real, and $\sin\xi=\sin\eta=0$ [cf.~Eqs.~(\ref{hbasis1}) and (\ref{hbasis2})].  That is,
\beq \label{etasignchange}
\xi, \eta=0~{\rm mod}~\pi\,.
\eeq
Since $\sin\xi=0$ in a real $\Phi_1$--$\Phi_2$ basis, it follows that $s_\beta e^{\pm i\xi}=s_\beta\cos\xi=\pm s_\beta$.  Thus, it is convenient to redefine $s_\beta\to s_\beta\,{\rm sgn}(\cos\xi)$, in which case the range of the parameter $\beta$ is extended to $-\half\pi\leq\beta\leq\half\pi$. 

In particular, $Y_3$, $Z_6$, $Z_7$, and 
\beq \label{twosigns}
\varepsilon\equiv e^{i\eta}=\cos\eta=\pm 1
\eeq
are real quantities that transform under a real orthogonal basis transformation,
$\Phi_i\to \mathcal{R}_{ij}\Phi_j$, as
\beq \label{etatransreal}
 [Y_3, Z_6, Z_7, \varepsilon]\to \det\mathcal{R}\, [Y_3, Z_6, Z_7, \varepsilon]\,,
\eeq
with $\det \mathcal{R}=\pm 1$.  In contrast, note that $Z_5$ is invariant under $\Phi_i\to \mathcal{R}_{ij}\Phi_j$.  
The quantity $\varepsilon=\pm 1$ accounts for the twofold nonuniqueness of the real Higgs basis.  
The parameter $Z_6$ plays a critical role in our analysis, and we shall henceforth assume that $Z_6\neq 0$. In light of \eq{etatransreal}, it follows that $\varepsilon Z_6$ is invariant under a real orthogonal basis transformation.

The neutral Higgs sector of the 2HDM consists of 
two CP-even neutral scalars, $h_1$ and $h_2$, with masses $m_1$ and $m_2$, respectively, and a CP-odd neutral scalar, $h_3=\sqrt{2}\Im\mathcal{H}_2^0$, with
mass $m_3\equiv m_A$, where
\beq
m_A^2=Y_2+\half(Z_3+Z_4-Z_5)v^2\,.
\eeq
The two neutral CP-even Higgs scalars are identified by diagonalizing a $2\times 2$ matrix, which is shown below~\cite{Connell:2023jqq} with respect to the basis $\{\sqrt{2}\Re\mathcal{H}_1^0-v\,,\sqrt{2}\Re\mathcal{H}_2^0\}$,
\beq \label{realmatrix33}
\mathcal{M}^2=\left( \begin{array}{ccc}
Z_1 v^2&\quad \varepsilon Z_6  v^2  \\
\varepsilon Z_6 v^2  &\quad m_A^2+Z_5v^2 \end{array}\right)\,,
\eeq
with corresponding mass-eigenstates
\beqa
h_1&=&\bigl(\sqrt{2}\,\Re \mathcal{H}_1^0-v\bigr)\cos\theta_{12}-\sqrt{2}\,\Re
 \mathcal{H}_2^0\sin\theta_{12}\,,\phantom{xxx}\label{hbasish}\\
h_2&=&\bigl(\sqrt{2}\,\Re  \mathcal{H}_1^0-v\bigr)\sin\theta_{12}+ \sqrt{2}\,\Re
 \mathcal{H}_2^0 \cos\theta_{12}\,,\phantom{xxx}\label{hbasisH}
 \eeqa
and masses $m_i\equiv m_{h_i}$ (where $m_1< m_2$) given by
 \beqa
 m^2_{1,2}&=&\half\Bigl[m_A^2+(Z_1+Z_5)v^2 \nonumber \\
 && \mp\sqrt{\bigl[m_A^2+(Z_5-Z_1)v^2\bigr]^2+4|Z_6|^2 v^4}\,\Bigr]\,,\label{cpevenmasses}
 \eeqa
in a notation where $m^2_1<m^2_2$.
The mixing angle $\theta_{12}$ is defined modulo $\pi$; by convention, we take $\cos\theta_{12}\geq 0$ and 
$-\half\pi\leq\theta_{12}\leq\half\pi$. 
In particular~\cite{Haber:2020wco},
\beqa
\sin \theta_{12}&=&\frac{\varepsilon Z_6 v^2}{\sqrt{(m_2^2-m_1^2)(m_2^2-Z_1 v^2)}}\,, \label{theta12s} \\[4pt]
\cos\theta_{12}&=&\sqrt{\frac{m_2^2-Z_1 v^2}{m_2^2-m_1^2}}\,.\label{theta12c}
\eeqa
That is, $\theta_{12}$ is basis-invariant and is thus independent of the twofold nonuniqueness of the real Higgs basis. 

In light of the LHC Higgs data~\cite{ATLAS:2022vkf,ATLAS:2025qxq,CMS:2022dwd,CMS:2026nce}, one of the two neutral CP-even Higgs bosons must resemble the SM Higgs boson. Indeed, if $\varphi\equiv \sqrt{2}\,\Re \mathcal{H}_1^0-v$ were a mass-eigenstate, its tree-level couplings would precisely match those of the SM Higgs boson.  The limiting case just described is known as the Higgs alignment limit~\cite{Ginzburg:2001wj,Gunion:2002zf,Craig:2012vn,Craig:2013hca,Asner:2013psa,Carena:2013ooa,Haber:2013mia,BhupalDev:2014bir} and corresponds to $\sin\theta_{12}=0$ if the lighter of the two CP-even Higgs bosons, $h_1$, is identified with the Higgs scalar discovered at the LHC~\cite{Bernon:2015qea}.   In particular, $h_1$ then resides entirely in the doublet $\mathcal{H}_1$ as indicated by \eq {hbasish}.

The notation introduced above differs from the standard notation employed in the 2HDM literature~\cite{Gunion:1989we,Branco:2011iw}, which was developed starting from a real $\Phi_1$--$\Phi_2$ basis where both the neutral components of $\Phi_1$ and $\Phi_2$ acquire a nonzero vev.
The diagonalization of the CP-even neutral scalar squared-mass matrix in this basis yields a mixing angle denoted by $\alpha$.
Neither $\alpha$ nor $\beta$ have a separate physical meaning in a general 2HDM~\cite{Haber:2006ue}.   However, after returning to the Higgs basis, one finds~\cite{Connell:2023jqq}
\beq \label{hH}
\begin{pmatrix} H\\ h\end{pmatrix}=\begin{pmatrix} \cbma & \,\,\, -\sbma \\
\sbma & \,\,\,\phantom{-}\cbma\end{pmatrix}\,\begin{pmatrix} \sqrt{2}\,\,{\rm Re}~\mathcal{H}_1^0-v \\
\varepsilon \sqrt{2}\,{\rm Re}~\mathcal{H}_2^0
\end{pmatrix}\,,
\eeq
where we have introduced the notation
$s_{\beta-\alpha}\equiv\sin(\beta-\alpha)$ and $c_{\beta-\alpha}\equiv\cos(\beta-\alpha)$. 
Comparing the equation above with Eqs.~(\ref{hbasish}) and (\ref{hbasisH}),
it follows that
\beq \label{bma}
\cos\theta_{12}=s_{\beta-\alpha}\,,\qquad
\sin\theta_{12}=-\varepsilon\,c_{\beta-\alpha}\,,
\eeq
or equivalently,
$\beta-\alpha=\varepsilon\,\theta_{12}+\tfrac{1}{2}\pi$.
Since $\theta_{12}$ is invariant with respect to real orthogonal basis transformations, it follows that $\sbma\geq 0$, whereas $c_{\beta-\alpha}$ changes sign under an improper real orthogonal basis change [cf.~\eq{etatransreal}].

Moreover, the neutral scalar mass eigenstates typically employed in the 2HDM literature~\cite{Gunion:1989we,Branco:2011iw} are related to the basis-invariant fields $h_i$ as follows~\cite{Haber:2022gsn,Connell:2023jqq}:
\beq \label{cpconservestates}
h=h_1\,,\qquad H=-\varepsilon\, h_2\,,\qquad A=\varepsilon\, h_3\,.
\eeq
Similarly, the charged Higgs scalar field employed in the 2HDM literature is given by
$H^\pm=\varepsilon\,\mathcal{H}_2^\pm$.  The corresponding charged Higgs mass is
\beq
m_{H^\pm}^2&=&Y_2+\tfrac{1}{2}Z_3 v^2
=m_A^2-\half v^2(Z_4-Z_5)\,.
\eeq

The scalar sector of the CP-conserving 2HDM is governed by the following eight independent parameters: $\{v,m_h,m_H,m_A,m_{H^\pm},Z_2,Z_3,Z_7\}$.  Note that the other $Z_i$ are fixed:
\beqa
Z_1 v^2&=& m_h^2\sbmaii+m_H^2\cbmaii\,,\label{zeeone}\\
Z_4 v^2&=& m_h^2\cbmaii+m_H^2\sbmaii+m_A^2-2m_{H^\pm}^2\,,\label{zeefour}\\
Z_5 v^2&=& m_h^2\cbmaii+m_H^2\sbmaii-m_A^2\,,\label{zeefive}\\
Z_6 v^2&=& (m_h^2-m_H^2)\cbma\sbma\,.\label{zeesix}
\eeqa

As $h$ is identified with the SM-like Higgs boson observed at the LHC, it then follows that $|\sin\theta_{12}|\ll 1$ (or equivalently, $|\cbma|\ll 1$).\footnote{Experimental evidence for $|\cbma|\ll 1$ is provided in Ref.~\cite{ATLAS:2024lyh}.} 
In light of \eq{theta12s}, the Higgs alignment limit 
can be achieved when $|Z_6|\ll 1$ and/or when $m_H\gg m_h$ (under the assumption that $Z_6$ lies in the perturbative regime).  In this paper, we propose to examine the 2HDM where $|Z_6|\lesssim\mathcal{O}(1)$ and $Y_2\gg v^2$, corresponding to the decoupling limit~\cite{Gunion:2002zf,Haber:2006ue}.  In this limit, $m_H\sim m_A\sim m_{H^\pm}\gg m_h$, or more precisely,
\beq
m_h^2&=& Z_1 v^2+\mathcal{O}\left(\frac{v^4}{m_A^2}\right)\,, \\
m_H^2&=& m_A^2+Z_5 v^2+\mathcal{O}\left(\frac{v^4}{m_A^2}\right)\,.
\eeq
Moreover, \eqs{theta12s}{bma} yield $Z_6\cbma<0$, with
\beq \label{cbmadecoup}
\cbma=-\frac{Z_6 v^2}{m_H^2}+\mathcal{O}\left(\frac{v^4}{m_A^4}\right)\,.
\eeq

\section{Decays of $H$ and $A$}
\label{DecaysH}

The Feynman rule for the $H V_\mu V_\nu$ coupling, where $VV=W^+W^-$ or $ZZ$, is given by $ig\lsub{HVV}g_{\mu\nu}$, with~\cite{Gunion:1989we}
\beq
g\lsub{HVV}=\frac{2m^2_V}{v} c_{\beta-\alpha}\,,\quad \text{for $V=W$, $Z$}.
\eeq
As expected, the $HVV$ couplings are suppressed close to the Higgs alignment limit, where $|\cbma|\ll 1$.  
There are no tree-level $AVV$ couplings.

Next, we record the coupling of $H$ to scalar final states.   The only possible kinematically allowed $H$ decay final states in the limit of new heavy Higgs boson masses $m_H\sim m_A\sim m_{H^\pm}$, where $m_H\gg v$, are $hh$ and $hhh$. The corresponding Feynman rules for the $Hhh$ and $Hhhh$ interaction vertices are given by $ig\lsub{Hhh}$ and $ig\lsub{Hhhh}$, respectively, where~\cite{Gunion:2002zf,Haber:2022gsn,Haberinprep}
\begin{widetext}
\beqa
g\lsub{Hhh}&=& {-3v}\bigl[Z_1\cbma\sbmaii
   +Z_{345}\cbma\left(\tfrac{1}{3}-\sbmaii\right)-Z_6\sbma(1-3\cbmaii)
    -Z_7\cbmaii\sbma\bigr]\,,\label{gHhhcoup} \\[5pt]
g\lsub{Hhhh}&=&    -3\bigl[Z_1\sbmaiii\cbma-Z_2\sbma\cbmaiii+Z_{345}\sbma\cbma(\cbmaii-\sbmaii) -Z_6\sbmaii(1-4\cbmaii)-Z_7\cbmaii(4\sbmaii-1)\bigr]\,,\nonumber
\\
&& \phantom{line} \label{gHhhhcoup}
\eeqa
\end{widetext}
with $Z_{345}\equiv Z_3+Z_4+Z_5$.
In the decoupling limit, where $|Z_6|\lesssim\mathcal{O}(1)$ and $|\cbma|\ll 1$ [as a result of \eq{cbmadecoup}], the scalar couplings above simplify to
$g\lsub{{Hhh}}\simeq  3v Z_6$ and $
g\lsub{{Hhhh}}\simeq 3Z_6$.
Using \eq{cbmadecoup}, it follows that
\beq \label{rat}
\frac{g_{HVV}}{g_{Hhh}} \simeq -\frac{2m_V^2}{3m_H^2}\,.
\eeq
The partial decay rates of $H\to VV$ and $H\to hh$ are directly correlated at leading order in $v^2/m_H^2$.
In particular, the corresponding partial decay rates of $H$ into $WW$, $ZZ$, $hh$, and $hhh$ are given by
\begin{widetext}
\beqa
\Gamma_{H\to VV}&=&c_V|g\lsub{HVV}|^2\frac{(m_H^4-4m_H^2 m_V^2+12m_V^4)\sqrt{m_H^2-4m_V^2}}{64\pi m_V^4m_H^2}\,, \quad \text{where $c_V=1~(1/2)$ for $V=W (Z)$}, \label{eq::2Z}\\
    \Gamma_{H\to hh} &=& \frac{1}{2!}\, |g_{Hhh}|^2\; \frac{1}{16 \pi m_H^2}\sqrt{m_H^2-4 m_h^2}\,,\label{eq::2h}
    \\
    \Gamma_{H\to hhh} &= & \frac{1}{3!}\, |g_{Hhhh}|^2\; \frac{1}{256 \pi^3m_H ^3}  \int_{4m_h^2}^{(m_H-m_h)^2} dx\,\lambda^{1/2}(x,m_H^2,m_h^2)\,\sqrt{1-\frac{4m_h^2}{x}}\,,
\label{eq::3h}
\eeqa
where $\lambda(a,b,c)\equiv a^2+b^2+c^2-2ab-2ac-2bc$, and we have made use of results given in Refs.~\cite{Gunion:1989we,Byckling} and Appendix~D of Ref.~\cite{Dreiner:2023yus}.  There are no decays of $A$ to scalar final states due to the assumed CP symmetry.
\end{widetext}

It is instructive to derive an approximate formula for $\Gamma(H\to hh)/\Gamma(H\to VV)$ in the limit where $|\cbma|\ll 1$ and $m_H\gg m_V, m_h$,
\beq \label{rathhVV}
\frac{\Gamma(H\to hh)}{\Gamma(H\to VV)}\simeq \frac{9}{2c_V}\left[1+\frac{2(3m_V^2-m_h^2)}{m_H^2}\right]\,.
\eeq
\Eq{rathhVV} yields asymptotic ratios of 9:2:1 for the decay rates of $H$ into the final states $hh$, $W^+ W^-$, $ZZ$, respectively.  The same asymptotic ratios were also obtained in Ref.~\cite{Li:2026kqk} in an extended scalar sector via the Higgs portal due to dimension-five scalar operators
(a phenomenon dubbed the busy Higgs mechanism). However, higher-dimensional operators are not needed in the 2HDM to reproduce the corresponding busy Higgs signal asymptotic ratios of Ref.~\cite{Li:2026kqk}.

Likewise, in the limit where $|\cbma|\ll 1$ and $m_H\gg m_h$,
\beqa
&&\frac{\Gamma(H\to hhh)}{\Gamma(H\to hh)}\simeq \frac{m_H^2}{96\pi^2 v^2}\left\{1-\frac{4m_h^2}{m_H^2}\left[1-3\ln\left(\frac{2m_h}{m_H}\right)\right]\right\},\nonumber \\
&& \phantom{line} \label{rathhhhh}
\eeqa
which is a factor of four smaller than the corresponding busy Higgs result with dimension-five scalar operators.

Next, we exhibit the couplings of $H$ and $A$
to quarks and leptons via the Yukawa Lagrangian~\cite{Gori:2017qwg,Connell:2023jqq}:
\beqa
&&\mathscr{L}_Y =
\frac{1}{v}\sum_{F=U,D,E} \overline F  \boldsymbol{M_F}\bigl[\varepsilon\, a_F\sbma-\cbma\bigr]FH\,, \\
&& \quad +\frac{i\varepsilon}{v}\left\{a_U\,\overline{U}\boldsymbol{M_U}\gamma\lsub{5}U-\sum_{F=D,E}a_F\,\overline{F}\boldsymbol{M_F}\gamma\lsub{5}F\right\}A\,,\phantom{xxx}
\label{YUK5}
\eeq
where $\varepsilon$ is defined in \eq{twosigns} and the $\boldsymbol{M_F}$ ($F=U,D,E$) are the corresponding diagonal 
up-type quark, down-type quark, and charged lepton 
mass matrices, respectively.  
The corresponding Feynman rules 
for the $Hf\bar{f}$ and $Af\bar{f}$ vertices
are given by $ig\lsub{Hf\bar{f}}$ and $ig\lsub{Af\bar{f}}$, where
\beqa 
  g\lsub{Hf\bar{f}}&=&\frac{m_f}{v}\Bigl(\varepsilon\, a_F\sbma-\cbma\Bigr)\,,\label{ghff} \\[4pt]
   g\lsub{Af\bar{f}}&=&\pm \frac{m_f}{v} i\varepsilon a_F\gamma\lsub{5}\,, \label{gaff} 
\eeqa
where the upper [lower] sign in \eq{gaff} corresponds to up-type [down-type] fermions, respectively.  
The partial decay rate of $H$ and $A$ into $t\bar{t}$ are given by
\beqa
\Gamma(H\to t\bar{t})&=&|g\lsub{Ht\bar{t}}|^2\,\frac{3m_H}{8\pi}\left(1-\frac{4m_t^2}{m_H^2}\right)^{3/2}\,, \label{GammaHtt} \\[4pt]
\Gamma(A\to t\bar{t})&=&|g\lsub{At\bar{t}}|^2\,\frac{3m_A}{8\pi}\left(1-\frac{4m_t^2}{m_A^2}\right)^{1/2}\,.
\eeqa
In the limit where $|\cbma|\ll 1$ and $m_H\gg m_t, m_h$,
\begin{align}
\frac{\Gamma(H\to hh)}{\Gamma(H\to t\bar{t})} \,\simeq\,\, &\frac{3v^4 Z_6^2}{4 m_H^2 m_t^2 (a_U-\varepsilon\cbma)^2} \nonumber \\[4pt]
& \qquad  \times\left[1+\frac{2(3m_t^2-m_h^2)}{m_H^2}\right],  \label{rathhtt}
\end{align}
where $Z_6$ is fixed by \eq{zeesix}.
 
Under an improper real orthogonal basis transformation, the couplings $g_{HVV}$, $g_{Hhh}$, $g_{Hhhh}$, $g_{Hf\bar{f}}$, and $g_{Af\bar{f}}$ 
\vskip 0.025in
\noindent
change sign, as expected in light of eqs.~(\ref{bma}) and (\ref{cpconservestates}).  Of course, the relevant physical quantities are the squares of these couplings, which are manifestly basis invariant.

Finally, we remark that one can set $\varepsilon=1$ thereby fixing the real Higgs basis, as no observable depends on this choice.\footnote{This choice is implicitly made in most treatments of the CP-conserving 2HDM in the literature.} 
Since $Z_6\cbma<0$ [cf.~\eq{cbmadecoup}], the signs of $Z_6$ and $\cbma$ are now physical but correlated. Moreover, due to the form of the $Hf\bar{f}$ coupling, it suffices to consider only positive values of $\cbma$ in a parameter space scan while allowing for values of $a_F$ of either sign.

\section{Phenomenological Analysis}
\label{pheno}

\begin{figure}[h!]
    \centering
    \includegraphics[width=1\linewidth]{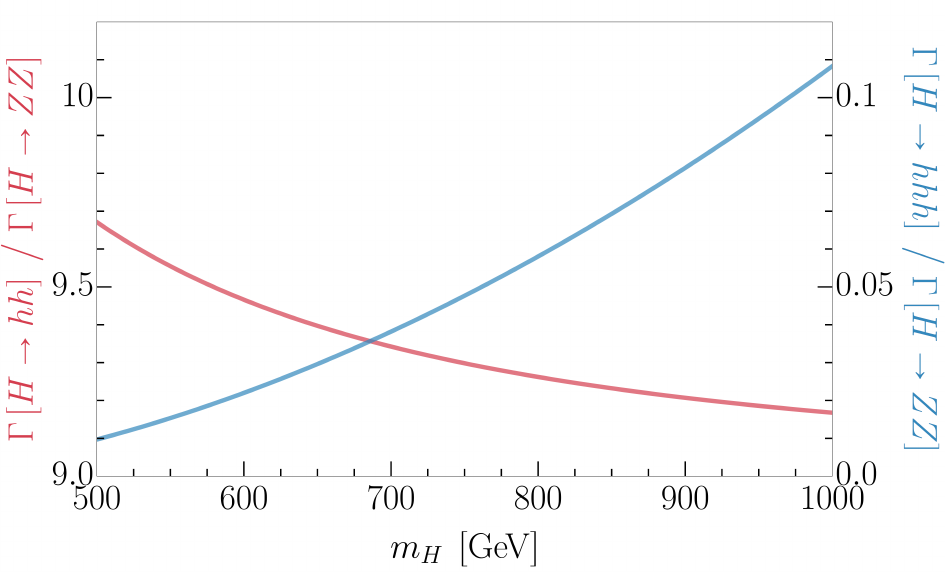}
    \caption{
    \justifying{
    2HDM predictions for $\Gamma(H\to hh)/\Gamma(H\to ZZ)$ (left axis, red) and $\Gamma(H\to hhh)/\Gamma(H\to ZZ)$ (right axis, blue) using the approximate expressions for the ratios of partial decay widths given in \eqs{rathhVV}{rathhhhh}.
    }
    }
    \label{fig:ratios}
\end{figure}

First, we examine the correlations among the decay widths (or equivalently, the branching ratios) given analytically in Sec.~\ref{DecaysH}.  The red curve in
Fig.~\ref{fig:ratios} depicts the ratio $\Gamma(H\to hh)/\Gamma(H\to ZZ)$, and the blue curve depicts the ratio $\Gamma(H\to hhh)/\Gamma(H\to ZZ)$, as a function of $m_H$.  
In the mass range of $500$\,GeV~$ < m_H < $~1\,TeV, \eqs{rathhVV}{rathhhhh} yield $9.65\gtrsim\Gamma(H\to hh)/\Gamma(H\to ZZ)\gtrsim 9.15$ and $0.01\lesssim\Gamma(H\to hhh)/\Gamma(H\to ZZ)\lesssim 0.1$. 
The approximate expressions used in obtaining
Fig.~\ref{fig:ratios} are reliable under the assumption that the self-coupling parameters $|Z_i|\lesssim\mathcal{O}(1)$ and $|\cbma|\ll 1$, which implies that the mass splittings among the heavy Higgs scalars are small [in light of \eqst{zeeone}{zeefive}].  Moreover, in the decoupling regime where $m_H\gg v$, \eq{zeesix} implies that $|Z_6|\sim (m_H^2/v^2)|\cbma|\gg |\cbma|$.  Consequently, the exact expressions for the ratios of partial decay rates 
exhibited in Fig.~\ref{fig:ratios} are insensitive to the specific choices for the other $Z_i$ due to the $\cbma$ suppressions in \eqs{gHhhcoup}{gHhhhcoup}.
Hence, the results of Fig.~\ref{fig:ratios} can be used in the future to test the 2HDM in case any of the processes mentioned above are observed.

Next, we consider the relevant constraints on $H$ production and decay in the decoupling regime.
In this setup, the vector boson fusion production cross section is suppressed. However, the second Higgs doublet $\mathcal H_2$ couples to top quarks [cf.~\eq{yukawas}], which induces gluon fusion production of $A$ and $H$ via a top quark loop. Using \eq{ghff} with $\varepsilon=1$, 
the gluon fusion cross section $gg\to H$ is given by
\begin{equation} \label{ggH}
    \sigma(gg\to H)=\bigl[a_U\sbma -\cbma\bigr]^2 \,\sigma(gg\to H_{\text{SM}})\,,
\end{equation}
where $H_\text{SM}$ is a hypothetical scalar with a mass of $m_H$ and with couplings to gauge bosons and fermions that coincide with those of the SM Higgs boson.

For $m_H>500$\,GeV, the dominant and most sensitive decay channels are $\Gamma_{H\to ZZ}$ [cf.~\eq{eq::2Z}], $\Gamma_{H\to hh}$ [cf.~\eq{eq::2h}], and $\Gamma_{H\to t\bar t}$.  In light of \eqs{ghff}{GammaHtt}, 
\begin{align} \label{Htt}
    \Gamma_{H\to t\bar t} &=\; \bigl[a_U\sbma -\cbma\bigr]^2 \,\Gamma_{H_\text{SM}\to t\bar t}\,.
\end{align}
The relevant experimental bounds can be found for $H$, $A\to t\bar t$ in Refs.~\cite{CMS:2025dzq,ATLAS:2024vxm}, for $H\to ZZ$ in Refs.~\cite{CMS:2026xbb,ATLAS:2020tlo} and for $H\to hh$ in Ref.~\cite{ATLAS:2023vdy}.
We have examined a benchmark point with $m_H=m_A=m_{H^\pm}=800$\,GeV, 
with $Z_1$, $Z_4$, $Z_5$, and $Z_6$ fixed via \eqst{zeeone}{zeesix}.
We have also set
$Z_2=Z_3=Z_7=1$, although our results are quite insensitive to the specific choices for 
$Z_2$, $Z_3$, and $Z_7$, 
as previously indicated.
For this benchmark point,
$t\bar t$ searches give a limit of $\bigl|a_U\sbma -\cbma\bigr|\lesssim 0.5$ for $H$ with $\Gamma_H /m_H = 1\%$\footnote{We checked that for the allowed parameter space, the ratio $\Gamma_H /m_H$ does not exceed 5\%.} and $\bigl|a_U\bigr|\lesssim0.4$ for $A$.
For the $H\to ZZ$ decay channel with $m_H=800$\,GeV, the currently observed LHC upper limit is $\sigma(gg\to H)\times\text{Br}[H\to ZZ]\approx0.01$\,pb, and for $H\to hh$ the upper limit is $\sigma(gg\to H)\times\text{Br}[H\to hh]\approx0.01$\,pb.\footnote{The searches for $t\bar t$ associated production of $A$ and $H$~\cite{CMS:2019rvj,ATLAS:2024jja} as well the charged Higgs in $tb$ associated production ($pp\to tbH^\pm$) with $H^\pm\to tb$~\cite{CMS:2019bfg,CMS:2022jqc,CMS:2021wlt,CMS:2025plw,ATLAS:2024hya,ATLAS:2021upq,ATLAS:2024rcu} do not yield competitive bounds.}

\begin{figure}
    \centering
\includegraphics[width=1\linewidth]{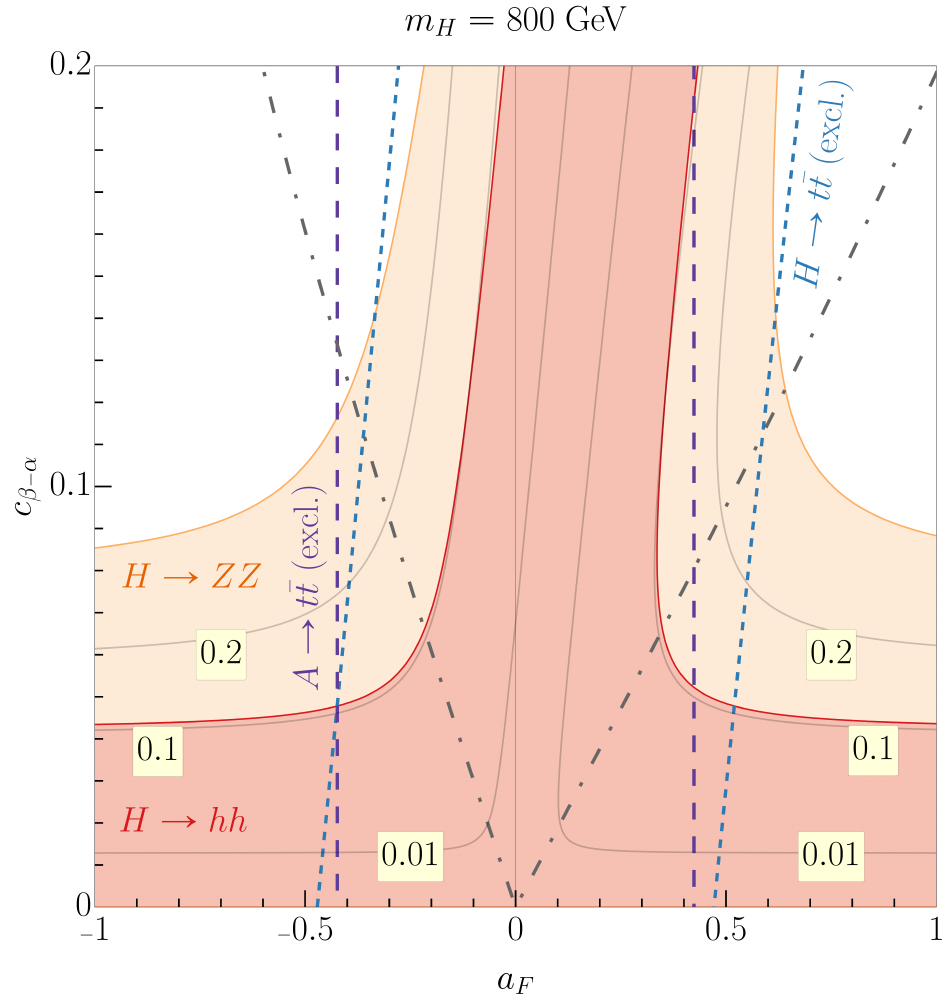}
    \caption{
    \justifying{
    Allowed regions in the $a_{F}$ vs.~$\cbma$ plane for $m_H\!=\!m_A\!=\! m_{H^\pm}\!=\!800$\,GeV, where $a_F\equiv a_U=a_D=a_E$. The contour lines show $\sigma(pp\to H \to hhh)$ in units of femtobarns for $\sqrt{s}=14$~TeV. The regions outside the dashed lines are excluded by $H\to t\bar t$ (blue) and $A\to t\bar t$ (violet) searches at the LHC.  In the region that lies inside the two dot-dashed lines, $H\to hh$ is the dominant decay mode [cf.~\eqs{rathhVV}{rathhtt} with $\varepsilon=1$].
    Note that in the decoupling limit with $|Z_6|\lesssim 1$, \eq{cbmadecoup} implies that $|\cbma|\lesssim 0.1$.}
    }
    \label{fig::limits}
\end{figure}
The results of our analysis are shown in Fig.~\ref{fig::limits}. Currently, for $m_H= 800$\,GeV, $H\to hh$ provides stronger constraints than $H\to ZZ$ over the whole parameter space.
Furthermore, in light of \eq{rathhhhh}, the cross section for prompt resonant tri-Higgs production $\sigma(pp\to H\to hhh)$ can be as large as $0.1$\,fb ($2$\,fb) for $\sqrt{s}=14$~TeV (100\,TeV)~\cite{Baglio:2015wcg}, which may be observable at the HL-LHC (100 TeV FCC-hh) with an integrated luminosity of 3000\,fb$^{-1}$ (20\,ab$^{-1}$). Focusing on the most sensitive tri-Higgs production channel, the $6b$ final state, with a $b$-jet efficiency of 
$\epsilon_b\approx0.85$ and a cut efficiency\footnote{The cut efficiency is derived from a cut-based analysis, while more recent studies based on deep learning techniques achieve an improvement by a factor of 6 resulting in approximately 18 events at the HL-LHC~\cite{Chiang:2025ecn}.} of $\epsilon_{\text{cut}}\approx0.1$ ($0.0115$)~\cite{Papaefstathiou:2023uum}, one can expect up to 3 (35) events. 

\begin{figure}
    \centering
    \includegraphics[width=1\linewidth]{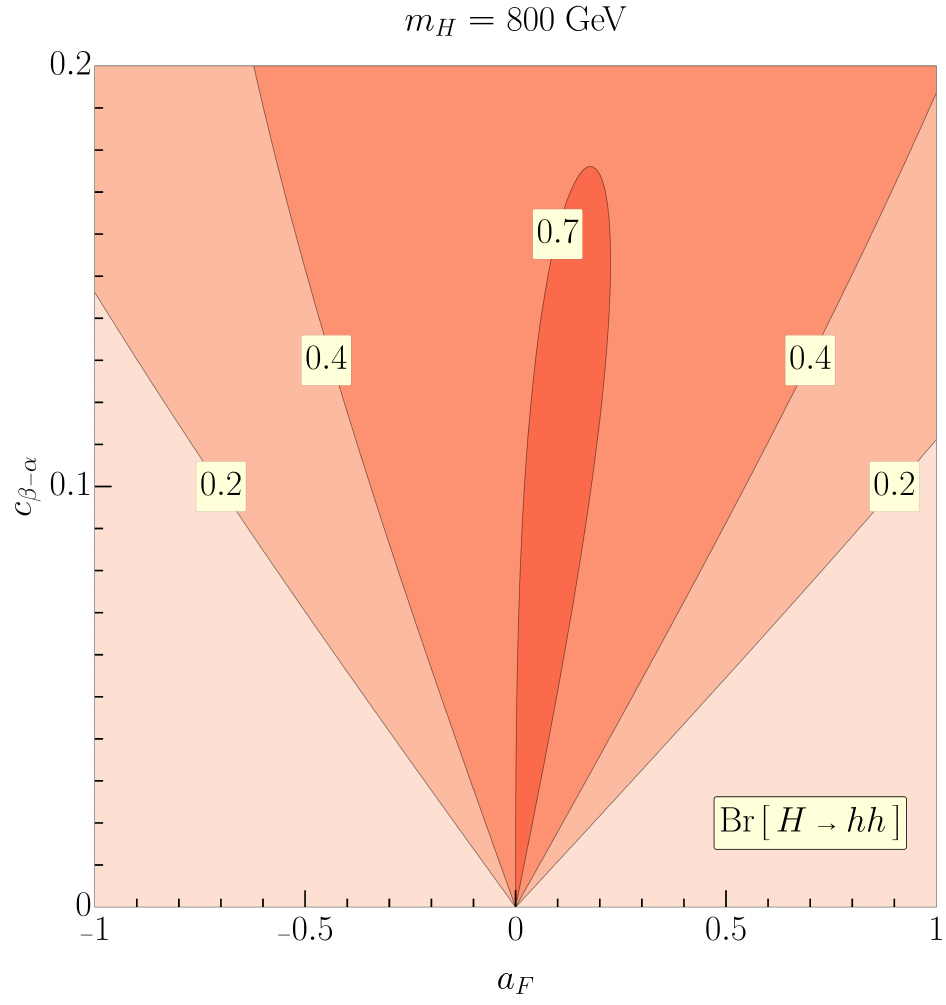}
    \caption{
    \justifying{The branching ratio of $H \to hh$ in the $a_{F}$ vs.~$\cbma$ plane for $m_H= m_A= m_{H^\pm}=800$\,GeV, where $a_F\equiv a_U=a_D=a_E$ and $\varepsilon=1$ [cf.~\eq{rathhtt}].  Contour lines corresponding to a branching ratio of 0.2, 0.4, and 0.7 are exhibited.}
    }
    \label{fig::brhh}
\end{figure}

In the allowed parameter space, the resonant decay $H\to hh$ can compete with $H\to t\bar t$ and can even become the dominant decay mode. This behavior, which is a consequence of \eq{rathhtt}, is highlighted within the region bounded by the dot-dashed lines in Fig.~\ref{fig::limits}.  Indeed,
the branching ratio of $H\to hh$ can exceed 70\% in some regions of the parameter space, as shown in Fig.~\ref{fig::brhh}.
As noted below \eq{Htt}, the LHC di-Higgs measurements constrain the production cross section to 
a value that is no larger than 0.01\,pb.  In particular, the di-Higgs production cross section is compatible with a relatively small top-quark alignment parameter $a_U$, which suppresses the $H\to t\bar t$ decay mode as indicated in \eq{rathhtt}. 

Note that the relations among the direct search channels discussed above hold at tree-level and are sensitive to loop corrections~\cite{Heinemeyer:2024qjp}. However, for $\mathcal{O}(1)$ self-couplings in the scalar potential, the loop effects are small. Furthermore, the constraints imposed by the LHC Higgs data are, in general,  satisfied close to the Higgs alignment limit. In particular, the deviation of the $h\to\gamma\gamma$ signal strength from the predicted value in the SM
is mainly sensitive to $Z_3$ (via the charged Higgs boson loop), which does not significantly affect the tree-level observables analyzed in this paper.  Finally, perturbativity 
constraints are easily satisfied
by choosing appropriately small mass splittings among $H$, $A$ and $H^\pm$.  

\section{Effects of CP-violation}

Although the analysis of Sec.~\ref{pheno} was carried out in the context of the CP-conserving 2HDM, our conclusions hold 
even in the case where additional sources of CP violation are present.\footnote{One must check that the parameters of the CP-violating 2HDM are not in conflict with the current experimental constraints~\cite{Roussy:2022cmp} on the electric dipole moment of the electron~\cite{Altmannshofer:2020shb}.} This can be seen as follows: In the CP-violating 2HDM, the three neutral scalar mass-eigenstates are states of indefinite CP, denoted by $h_1$, $h_2$, and $h_3$.  We adopt the convention where $m_1 < m_2$, $m_3$ (where $m_k\equiv m_{h_k}$), and assume that $h_1$ can be identified with the SM-like Higgs boson $h$ observed at the LHC. Indeed, in the Higgs alignment limit, $h_1$ is approximately CP-even, whereas $h_2$ and $h_3$ are not CP eigenstates.  The three neutral scalar mass-eigenstates are obtained by diagonalizing a $3\times 3$ squared-mass matrix, and as shown in Refs.~\cite{Haber:2006ue,Boto:2020wyf}, the diagonalization procedure yields two physical mixing angles, denoted by $\theta_{12}$ and $\theta_{13}$.  

In the decoupling limit where $m_2\sim m_3\sim m_{H^\pm}\gg v$, \eq{cbmadecoup} is replaced by~\cite{Haber:2022gsn}:
\beqa
\sin\theta_{12}&=&\frac{v^2\Re(Z_6 e^{-i\eta})}{m_2^2}+\mathcal{O}\left(\frac{v^4}{m_2^4}\right)\,,  \label{decoup1}\\
\sin\theta_{13}&=&-\frac{v^2\Im(Z_6 e^{-i\eta})}{m_3^2}+\mathcal{O}\left(\frac{v^4}{m_3^4}\right)\,.\label{decoup2}
\eeqa
The couplings of $h_k$ to $VV$, $hh$, and $hhh$ in the decoupling limit are given at leading order by
\beqa
g\lsub{h_kVV}&\simeq &\frac{2m_V^2}{v}\sin\theta_{1k}\,,\quad \text{for $k=2,3$},\\
g_{h_2 hh}&\simeq & -3v\Re(Z_6 e^{-i\eta})\,,\\
g_{h_3 hh}&\simeq & \phantom{-}3v\Im(Z_6 e^{-i\eta})\,,\\
g_{h_2 hhh}&\simeq & -3\Re(Z_6 e^{-i\eta})\,,\\
g_{h_3 hhh}&\simeq & \phantom{-} 3\Im(Z_6 e^{-i\eta})\,.
\eeqa
In particular, in light of \eqs{decoup1}{decoup2}, we obtain
\beq
\frac{g_{h_k VV}}{g_{h_k hh}}\simeq -\frac{2m_V^2}{3m_k^2}\,,
\quad \text{for $k=2,3$},
\eeq
which generalizes the result of \eq{rat}.

Similarly, the couplings of $h_2$ and $h_3$ to fermions are obtained from the Yukawa Lagrangian at leading order in the decoupling limit, 
\beqa\mathscr{L}_Y&=&-\frac{1}{v}\Biggl\{\overline{U}\boldsymbol{M_U}\bigl(a^U P_R+a^{U*}P_L\bigr) \nonumber \\
&& \quad +\sum_{F=D,E}\overline{F}\boldsymbol{M_F}\bigl(a^{F*}P_R+a^F P_L\bigr)\Biggr\}h_2\phantom{xxxx} \nonumber \\[4pt]
&&
-\frac{i}{v}\Biggl\{\overline{U}\boldsymbol{M_U}\bigl(-a^U P_R+a^{U*}P_L\bigr) \nonumber \\
&& \quad +\sum_{F=D,E}\overline{F}\boldsymbol{M_F}\bigl(a^{F*}P_R-a^F P_L\bigr)\Biggr\}h_3\,,
\eeqa
where $P_{R,L}\equiv\half(1\pm\gamma\lsub{5})$ and the flavor-alignment parameters $a^F$ ($F=U,D,E$) may be complex.
However, the magnitude of the corresponding couplings of $h_2$ and $h_3$ to fermion pairs will not differ significantly from the corresponding CP-conserving case.

Therefore in the CP-violating 2HDM, the analysis of Sec.~\ref{pheno} can be applied to resonant di-Higgs and tri-Higgs production via resonant $h_2$ and $h_3$ production and decay,\footnote{The interplay between the Higgs alignment limit of the 2HDM and CP-violating observables has been studied via non-resonant cascade di-Higgs and tri-Higgs production in Refs.~\cite{Low:2020iua,Chen:2022vac}.} where $|Z_6|$ is replaced by the basis-independent scalar self-couplings $\Re(Z_6 e^{-i\eta})$
and $\Im(Z_6 e^{-i\eta})$, respectively. 
Note that the decay $h_2\to \gamma\gamma$ [$h_3\to\gamma\gamma$] 
is mediated at one loop primarily via the couplings of $h_2$ [$h_3$] to $W^+W^-$, $t\bar t$, and $H^+ H^-$,
where the latter is proportional to
$\Re(Z_7 e^{-i\eta})$ [$\Im(Z_7 e^{-i\eta})$], respectively~\cite{Banik:2024ftv,Banik:2024ugs}.
However, the corresponding $h_2$ and $h_3$ decay widths are negligible in the region of parameter space considered in this work.

\section{Conclusions}
\label{conclude}

In this paper, we have derived simple relations among the decay widths of the heavier CP-even Higgs boson of the 2HDM (denoted by $H$) into $hh$, $ZZ$, $WW$ and $hhh$ that hold for $m_H\gg v$, i.e.~close to the Higgs alignment limit in the decoupling regime. In particular, we find that $9.65\gtrsim \Gamma(H\to hh)/\Gamma(H\to ZZ)\gtrsim 9.15$ and $0.01\lesssim\Gamma(H\to hhh)/\Gamma(H\to ZZ)\lesssim 0.1$ when $500$\,GeV~$<m_H<$~1\,TeV. These relations can be used as simple tests of the 2HDM if a new scalar particle is discovered in one of these decay modes. Furthermore, by taking into account the current experimental LHC bounds on $H\to ZZ$, $H\to hh$, $H\to t\bar t$ and $A\to t\bar t$, we find that a significant region of the parameter space exists where $H\to hh$ is a dominant decay mode and serves as the 
leading probe for the discovery of a heavy scalar narrow resonance at current and future hadron colliders.
Furthermore, $\sigma(pp\to H\to hhh)$ can be as large as $0.1$\,fb at $\sqrt{s}=14$~TeV, thereby suggesting that resonant prompt tri-Higgs production in a 2HDM might be observable at the HL-LHC (with better prospects at a future FCC-hh collider at $\sqrt{s}=100$~TeV). 

\acknowledgements
G.C. and H.E.H. thank Zhen Liu for an illuminating discussion of his recent work and the relation of our result to the busy Higgs signal introduced in Ref.~\cite{Li:2026kqk} subsequent to our paper.
A.C. and G.C.~acknowledge support from a professorship grant from the Swiss National Science Foundation (No.\ PP00P21\_76884).
H.E.H. is supported in part by the U.S.~Department of Energy Grant
No.~\uppercase{DE-SC}0010107.

\appendix

\bibliographystyle{utphys}
\bibliography{bibliography}
\end{document}